# Mirrored strain-balanced quantum well concentrator cells in the radiative limit

J. P. Connolly[1], M. F. Führer[1], D.C. Johnson[1], I.M. Ballard[1], K.W.J. Barnham[1], M. Mazzer[1], T.N.D Tibbits[1],
J.S. Roberts[2], G. Hill[2], C. Calder[2]

[1]Experimental Solid State Physics, Imperial College London, London, SW7 2BW, UK
[2]EPSRC III-V Facility, University of Sheffield, Mappin Street, Sheffield S1 3JD, UK

## ABSTRACT

Strain-balanced Quantum Well Solar Cells [1] (SB-QWSCs [2]) are radiatively dominated at concentrator current levels [3,4]. Incorporating back surface mirrors on the back of the doped substrate leads to a reduction in this radiative recombination current while the non-radiative dark current at lower bias remains unchanged [5]. This work extends a previous model of radiative dark currents [6,7,8] to cells where the substrate is not a perfectly absorbing sink for emitted luminescence by including multiple passes of photons emitted by radiative recombination. The luminescence in such mirrored structures is reduced and the radiative dark current lower. The structure is suited to concentrator applications because it corresponds to the high efficiency limit where losses due to light re-emission are limited to the solid angle for absorption of solar radiation [8].

## 1. Introduction

The strain balanced quantum well solar cell (SB-QWSC [1,2]) is a *p-i-n* structure with quantum wells encased between higher gap barriers in the field bearing intrinsic region. In the InGaAs/GaAsP system on GaAs substrates the compressive strain in the InGaAs quantum wells is balanced by tensile strain in the GaAsP barriers allowing strained designs without dislocations. This allows free specification of cell bandgaps beyond lattice matched materials. The wells extend the absorption spectrum while the flat band potential remains determined by the Fermi levels in the GaAs *p* and *n* charge neutral layers. Although the wells increase recombination rates this design tends to benefit from high open circuit voltage because of the larger built in voltage.

This paper examines additional high forward bias features of this design which are attractive for concentrator applications:
1. The inherently radiatively dominated behaviour of the SB-QWSC at concentrator current levels [2].
2. The ideal Shockley recombination is much smaller than the radiative due to the higher band gap of the homogeneous parts of the cell [1, 2].
3. The bulk regions of the cell are transparent to most of the radiative recombination.
4. The relatively low absorption of the wells leads to a significant increase in photocurrent brought about by including back surface mirrors.

The two effects of photocurrent increase and radiative behaviour dominating provide two mechanisms increasing the operating voltage, fill factor and efficiency.

The analysis of the radiative current in forward bias shows further interesting points. The first is the reduction of luminescence towards the substrate and consequent reduction in the dominant radiative recombination current. We will see that the reduced radiative recombination in mirrored QWSCs can restore ideal Shockley behaviour. This yields a cell with a fundamental band-edge determined by the wells but a recombination determined by the high bandgap charge neutral layers [1].

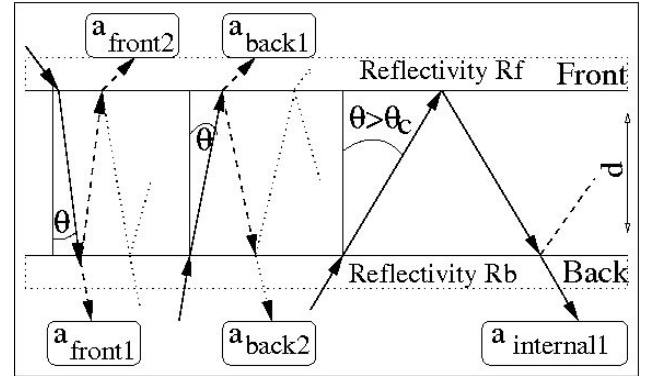

**Figure 1**. Light absorption path series in the cell of thickness *d* from the front ($a_{front}$) within the front surface critical angle $\theta c$ and from the back within ($a_{back}$) and outside $\theta c$ ($a_{int}$).

Finally, a study of the electroluminescence in low well number devices has shown higher carrier temperatures in the wells than the bulk regions [14]. This provides a mechanism for thermodynamic consistency of our earlier observation of reduced quasi-Fermi levels in the wells [6].

## 2. Theory

The quantum efficiency (QE) of the SB-QWSC is calculated by analytically solving transport equations for charge neutral layers and within the depletion approximation for the depletion layer incorporating the wells [9] with modifications for strained materials [3,10]. The absorption coefficient is taken from published data for bulk regions and in the wells is calculated by solving the Schrödinger equation in the effective mass approximation [9]. The integral of the product of the QE and the incident flux density gives the short circuit current $I_{SC}$.

The dark current is the sum of ideal Shockley $I_S$, Shockley-Read-Hall non radiative $I_{SRH}$, and radiative $I_{RAD}$ currents [3]. Assuming superposition of light and dark currents (verified in Ref. 11), the current under illumination $I_L$ is then

$$I_L(V) = I_{SC} - (I_S(V) + I_{SRH}(V) + I_{RAD}(V)) \quad (1)$$

We now extend the radiative current for a perfect back surface absorber [6] to describe cells with back surface mirrors. The net radiative current $I_{RAD}$ is determined by the generalised Planck equation for cell refractive index *n*, photon energy *E* and quasi-Fermi level separation or chemical potential $\Delta\phi$ as

$$I_{RAD} = \int_0^\infty \left( q \frac{2n^2}{h^3 c^2} \left( \frac{E^2}{e^{(E-q\Delta\phi)/kT} - 1} \right) \int_S a(E,\theta) dS \right) dE \quad (2)$$

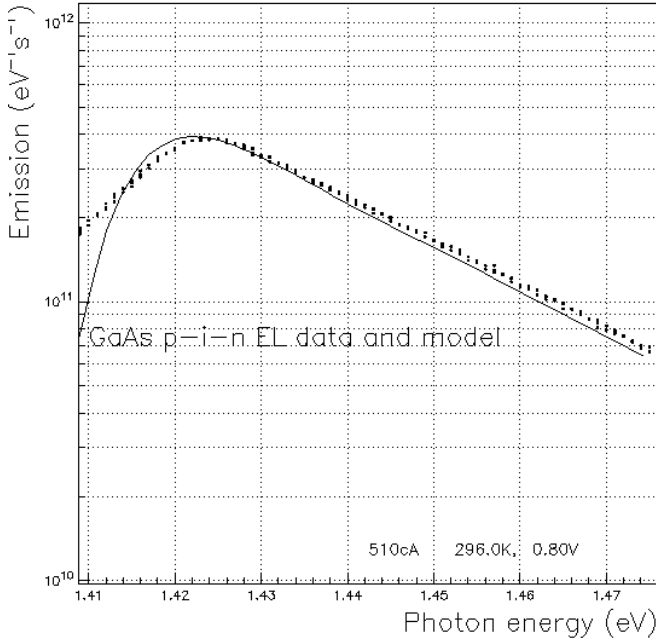

Figure 2. EL data (dots) in absolute units and theory (line) of a GaAs p-i-n showing 296K for measurement temperature 295K

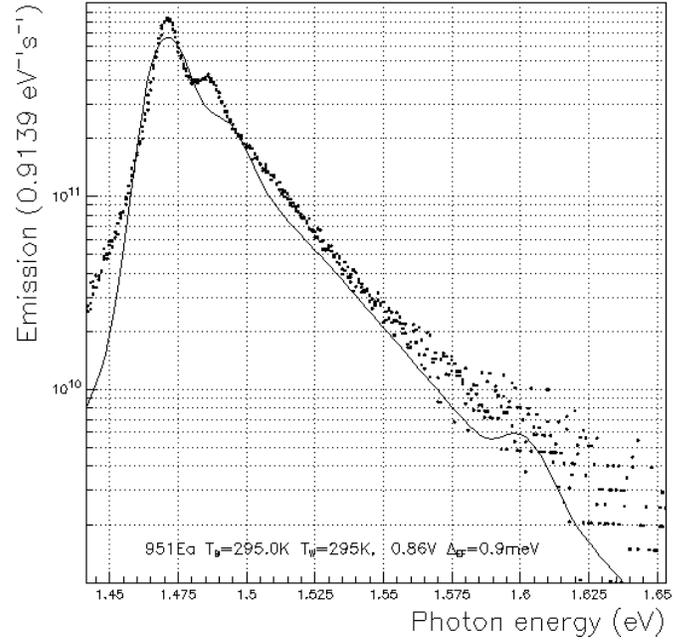

Figure 3. EL data (dots) in absolute units and theory (line) of a 50 well GaAs/AlGaAs MQW showing accurate scaling (0.91 vs. exact value 1), temperature and negligible Δø of 0.9meV.

where $a(E,\theta,s)$ is the absorptivity of the cell at point $s$ on the surface of the cell at internal angle $\theta$ to the normal.

The absorptivity $a(E,\theta,s)$ is given by the sum of possible cell emission paths in terms of absorption coefficient $\alpha(E)$, and front and back surface areas and reflectivities $A_{Front}$, $R_F$, $A_{Back}$ and $R_B$ as illustrated in figure 1.

Light incident on the front of the cell within the critical angle may pass straight through the cell of thickness $d$ and be emitted through the back surface (path $a_{front1}$), or be internally reflected and partially re-emitted from the front surface (path $a_{front2}$) and so on to infinity. Free carrier absorption in the doped substrates considered is low in this wavelength range [12] and is neglected.

The total absorptivity from the front for $\theta < \theta c$ takes the form of two geometric sums as follows

$$a_{front}(E,\theta,s) = (1-Rf)\begin{pmatrix} 1-(1-Rb)\sum_{n=1}^{\infty} e^{-(2n-1)\alpha d \sec(\theta)}(RfRb)^{n-1} \\ -(1-Rf)\sum_{n=1}^{\infty} Rb e^{-(2n)\alpha d \sec(\theta)}(RfRb)^{n-1} \end{pmatrix} \quad (3)$$

We assume a constant refractive index at the back surface of the cell which we define as the interface between the cell and the doped GaAs substrate. The critical angle at the back surface is then $\theta c(back) = \pi/2$. For light absorbed through the back of the cell the picture is reversed except that the critical angle is $\theta c(front) < \pi/2$ giving one series $a_{back}$ for angles within the critical angle and one series $a_{internal}$ for angles greater than $\theta c(back)$ with total internal reflection.

The absorptivity $a_{bac}$ from the back for $\theta < \theta c$ is given by inversing Rf and Rb in equation (3), and $a_{internal}$ given by setting Rf to 1 in the expression for $a_{back}$. Omitting details for brevity the infinite sums for absorptivity are

$$a_{front}(E,\theta,s) = \frac{(1-e^{\alpha d \sec(\theta)})(Rf-1)(e^{\alpha d \sec(\theta)}+Rb)}{e^{2\alpha d \sec(\theta)}-RbRf}$$

$$a_{back}(E,\theta,s) = \frac{(1-e^{\alpha d \sec(\theta)})(Rb-1)(e^{\alpha d \sec(\theta)}+Rf)}{e^{2\alpha d \sec(\theta)}-RbRf} \quad (4)$$

$$a_{internal}(E,\theta,s) = \frac{e^{\alpha d \sec(\theta)}(Rb-1)(e^{\alpha d \sec(\theta)}-e^{-\alpha d \sec(\theta)})}{Rb - e^{2\alpha d \sec(\theta)}}$$

and the total absorptivity integral $Ad$ over surface and solid angle of equation (2) in terms of surfaces areas $A_F$, $A_B$ is then calculated separately for well and bulk as

$$Ad = \int_S a(E,\theta)dS$$
$$= 2\pi A_F \int_0^{\theta_C} a_{front} \cos(\theta)d\theta \quad (5)$$
$$+ 2\pi A_B \left( \int_0^{\theta_C} a_{back} \cos(\theta)d\theta + \int_{\theta_C}^{\pi/2} a_{internal} \cos(\theta)d\theta \right)$$

The radiative rate at energy E (or luminescence) is then

$$L_{RAD} = q\frac{2n^2 E^2}{h^3 c^2}\left[\left(\frac{Ad_{Bulk}}{e^{(E-q\Delta\phi_{Bulk})/kT_{Bulk}}-1}\right) + \left(\frac{Ad_{Well}}{e^{(E-q\Delta\phi_{Well})/kT_{Well}}-1}\right)\right] \quad (6)$$

where we have explicitly separated well and bulk contributions, introducing bulk and well carrier temperatures $T_{Bulk}$, $T_{Well}$, quasi-Fermi level separations $\Delta\phi_{Bulk}$, $\Delta\phi_{Well}$, and absorptivities $a_{Bulk}$, $a_{Well}$. These determine $Ad_{Bulk}$ and $Ad_{Well}$ as per equation (5). $I_{RAD}$ is then given by equation (2).

### 3. Results and discussion

We examine two examples to validate the analysis. Figure 2 shows the electroluminescence (EL) model for a bulk p-i-n solar cell previously reported in ref. [3] where sample details and QE modeling are presented. The EL peak is slightly offset and the below gap EL decay is weaker in the data, which is ascribed to inaccurate low absorption bulk GaAs data. However the error in the integral and hence $I_{RAD}$ is negligible. Figure 3 shows similar results for published data [9] in absolute units for a 50 well lattice matched GaAs/AlGaAs MQW. In this case the well exciton features are imperfectly modeled but the fitting model described further on correctly estimates the calibration factor, the well temperature is found equal to the bulk, and a negligible Δø of 0.9meV is estimated.

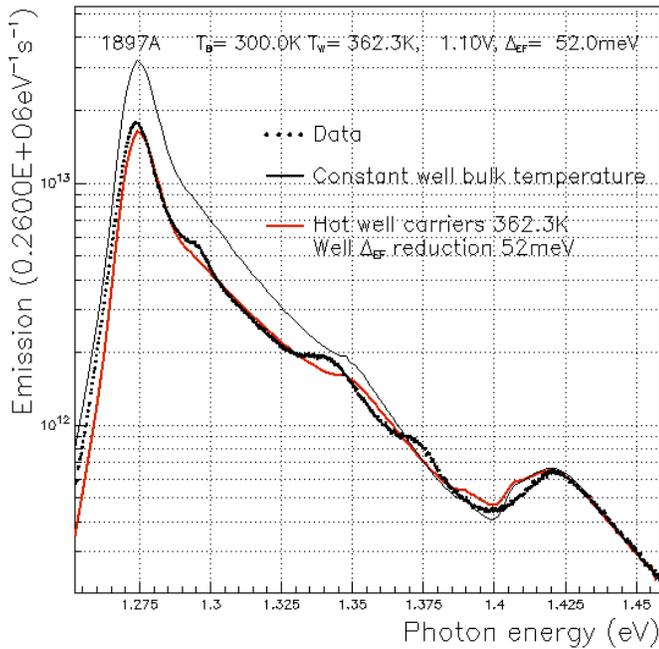

Figure 4 Electroluminescence modeling of a ten well QWSC showing high well carrier temperature signature and suppressed quasi-Fermi levels in the wells.

We now progress to a ten well SB-QWSC, experimental and sample details of which are reviewed in previous work [13]. This work analysed the EL assuming equal well and bulk carrier temperatures at low bias around 0.9V. This found evidence of suppressed $\Delta\phi_{Well}$ compared to $\Delta\phi_{Bulk}$ (equal to applied bias) but essentially the same carrier temperatures in well and bulk.

More recent numerical analysis of EL at higher bias has shown however the in this bias regime above a volt there is evidence in the shape of the EL of higher well carrier temperature in the wells than in the bulk [14]. This is visible in terms of a difference in the slope of the EL in the regions dominated respectively by bulk and well. The analysis of Ref. 14 has shown that using measured data to estimate $Ad_{Bulk}$ and $Ad_{Well}$ and using a least squares fit to extract $T_{Well}$ correctly finds 300K for the bulk but a higher temperature in the wells by some tens of degrees Kelvin.

In this work we extend this analysis in order to generalise modeling of $I_{RAD}$. We assume that the bulk temperature is equal to the ambient and that $\Delta\phi_{Bulk}$ is equal to the applied bias because of correct modeling of bulk samples (Fig. 2).

Since the data is not yet in absolute units for experimental reasons we are left with three unknowns which are the calibration factor, $T_{Well}$ and $\Delta\phi_{Well}$.

By numerical analysis of equation 6 we find that a single minimum is defined for the three unknowns essentially because they are not linearly related. A least squares method, which integrates the model described above, yields the values of all parameters for a given EL data file in arbitrary units. It has been verified by tests with known synthetic data sets as well as previously published data in absolute units (figs. 1,2).

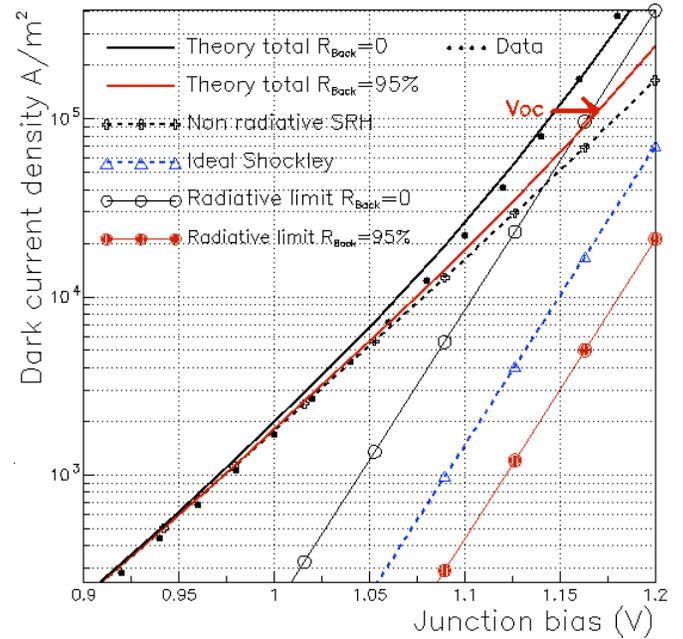

Figure 5 Dark current without (upper black) and with (red) back surface mirrors of 95% showing decreased dark current and increased Voc for efficiency quoted in the main text.

Applying this methodology we find in agreement with the method of ref. [14] that the well luminescence is over-estimated and the variation of EL with energy disagrees with the data at high bias if we assume a constant well and barrier temperature. We find a good fit however with a well temperature increase of 62K and total quasi-Fermi level suppression of 52meV for this data taken at 1.1V

Two complementary methods have therefore found temperature increases in the well. However the magnitude of the temperature increase depends on whether the quasi-Fermi level separation is allowed to vary. This work has in addition shown that a varying suppression of $\Delta\phi_{Well}$ is found which combined with the variation of $T_{Well}$ ensures an ideality of 1. From the point of view of dark current modeling, we find we can use a constant combination of these two parameters since the dark current is sensitive to the integral of the EL and not to the wavelength dependence. The following dark current modeling therefore uses zero $\Delta\phi_{Well}$ and $T_{Well}$ equal to the bulk temperature as found at low bias in earlier work. This varies with ideality 1 as does the exactly determined bias dependent proper evaluation of $T_{Well}$ and $\Delta\phi_{Well}$ which is difficult to perform across the entire bias range and is work in progress.

We conclude that within this important approximation the dark current remains correctly modeled. Figure 5 shows calculated dark current theory and data in black for a non mirrored 50 well cell (and corresponding $I_{RAD}$ in empty circles) for which details are given in Ref. 11. The reduced modeled dark current with a 95% back surface mirror is shown in red ($I_{RAD}$ full circles). The ideal Shockley current $I_S$ which is calculated from transport parameters evaluated by fitting the bulk QE [3] stays constant with and without back mirrors. The non radiative $I_{SRH}$ current is determined by a single fitting parameter which is the non radiative lifetime in the depletion region and also remains unchanged with and without mirrors.

The calculated efficiency is 27.0% for the non mirrored cell under AM1.5D with 300x concentration. This is comparable with previous measured concentrator results [11].

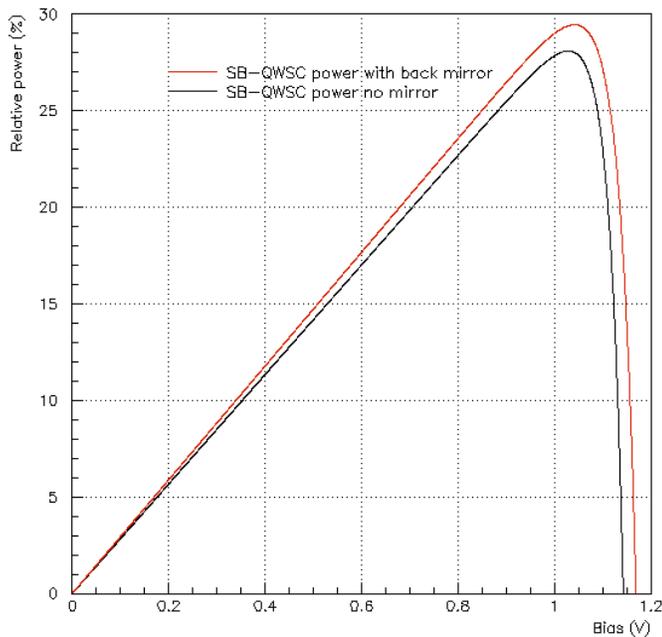

Figure 6 Power output divided by solar flux for cells with and without back mirrors - the peaks are the efficiencies.

This increases to 28.4% efficiency for the modeled mirror cell with a Voc increase of 10mV which, if achievable, would be a single-junction record.

## 4. Conclusions

We have shown increases in efficiency from 27% to a modeled 28.4% for a 50 well SB-QWSC under 300xAM1.5D concentration by coating the doped substrate with a mirror. The analysis shows that this is due to dark current reduction coupled with increased photocurrent.

The modeling involved has shown new promising aspects. The first and least understood that we have seen is the signature of hot carriers verified by two methods. Indications are that this effect not seen in lattice matched systems appears at relatively high voltages and in strained systems. In this case the band-gap in the barriers is higher than in the neutral layers which may reduce ideal Shockley behaviour. It also suggests an interplay between high carrier densities as seen in hot carrier effects in lasers, and spatial confinement or density of states modulation of phonon populations.

The counter-intuitive concept of hot carriers in the steady state requires non equilibrium dynamics. One way this may occur is if position dependent temperature is offset by position dependent carrier populations in the steady state. This is a consequence of terms in transport equations dependent on thermal gradients being counterbalanced by terms dependent on carrier density [15]. Since we see evidence of position dependent carrier densities and varying temperatures there is no a priori reason for inconsistency. This is a promising avenue for further investigation.

The second promising phenomenon we have shown is the suitability of QWSCs for concentrator applications due to the inherent radiative tendency of these devices at high current levels. We have shown previously that this is not the case for bulk designs which tend towards the less favourable ideal Shockley limit. Therefore at high currents SB-QWSCs tend towards the classic high efficiency limit.

We have shown that SB-QWSCs with good material quality have a promising feature at high concentration where they operate in the radiative limit. They are suited to the technique of emission angle reduction or equivalently photon recycling. By the addition of a mirror at the back surface we have shown that the NET emission towards the substrate is suppressed.
That is, photons emitted towards the rear tends to be trapped in the cell or recycled. This is inherent in the Planck formalism we have applied here and is expressed in terms of infinite sum of possible emission paths, or equivalently possible absorption pathways.

These considerations allow us to postulate a simple means of improving a cell leading to a single-junction cell of 28.4%.